\documentclass[preprint,preprintnumbers,amsmath,floatfix,balancelastpage,nofootinbib,prb]{revtex4-1}
\usepackage{graphicx, braket, pgfplots, fancyhdr, natbib, multirow,relsize}

\pgfplotsset{compat=newest} 
\pgfplotsset{plot coordinates/math parser=false} 
\newlength\figureheight 
\newlength\figurewidth 

\usepackage[margin=0.7in]{geometry} 
\newcommand*\rfrac[2]{{}^{#1}\!/_{#2}} 
\usepackage[colorlinks=true,citecolor=blue]{hyperref} 
\usepackage{cleveref} 

\fancypagestyle{plain}{%
  \fancyhf{}
  \fancyfoot[C]{\thepage}%
}

\begin{document}


\title{\textbf{The Effect of Next-Nearest Neighbour Hopping in the One, Two, and Three Dimensional Holstein Model}}
\author{Carl J. Chandler}
\author{Christian Prosko}
\author{F. Marsiglio}
\email{fm3@ualberta.ca}
\affiliation{Department of Physics, University of Alberta, Edmonton, Alberta, Canada T6G 2E1}
\date[Dated ]{\today}


\begin{abstract}
\begin{footnotesize}
Allowing a single electron to hop to next-nearest neighbours (NNN) in addition to the closest atomic sites in the Holstein model, a modified Trugman method is applied to exactly calculate the effect on the polaronic effective mass in one, two, and three dimensions, building on the previous study of the one-dimensional NNN Holstein model. We also present perturbative calculations and a heuristic scaling factor for the coupling strength and ion frequency to nearly map 
the NNN Holstein model back onto the original Holstein model. When account is taken of the modified electronic bandwidth near the
electron energy, we find that including NNN hopping effectively {\it increases} the polaron effective mass.

\end{footnotesize}
\end{abstract}
\maketitle

\thispagestyle{plain} 

\noindent {\bf INTRODUCTION}
\vskip1.0cm

In the realm of BCS theory, it is well known that electron-phonon interactions in solid materials are integral to the emergence of superconductivity, as they are responsible for the effective attraction that leads to the formation of Cooper 
pairs.\cite{bardeen1957} On the other hand, the 
importance of electron-phonon interactions in high temperature superconductivity is not yet clear.\cite{alexandrov2011}
Since polarons are simply quasiparticles consisting of electrons dressed with the net effect of these electron-phonon interactions, 
it is important to understand this basic building block to fully understand conventional superconductivity, and possible extensions
to nonconventional superconductors.\cite{hirsch15} To this end, the problem 
of a single electron in the conduction band of a crystal lattice has been extensively studied.\cite{alexandrov07} Specifically, a
numerically exact algorithm for solving the Holstein model with tight-binding electron bands in the thermodynamic limit was formulated in
Ref. \onlinecite{bonca99}, and now that problem is effectively solved. Several extensions were subsequently
reported, including ones to better manage disparate electron ($t$) and phonon ($\omega_E$) energy scales (in particular, 
$\omega_E <<t$),\cite{li10,alvermann10}, higher dimensionality,\cite{ku02,li12,davenport12} extended interaction range \cite{bonca01,holsteinchandler14} and inclusion of next-nearest neighbour (NNN) single-particle hopping amplitude.\cite{chakraborty2011}
In this last study it was found that including NNN hopping in the one dimensional Holstein model altered significantly the electron's 
effective mass in strong coupling. 

The purpose of this paper is to follow up on this study. Thus far studies of polaron properties within the Holstein model have revealed
that the effective mass becomes very large with rather modest electron-phonon coupling strength. This is incompatible with
experiment, specifically with the evidence that some conventional superconductors have a large electron-phonon coupling
strength, and yet show almost no sign of single-electron polaronic behaviour in the normal state.\cite{gladstone69}
Chakraborty \textit{et al.} \cite{chakraborty2011} found that including NNN hopping in the one-dimensional Holstein model could
decrease the polaron effective mass significantly, particularly at strong coupling. This is potentially very important since this is a means for lowering the polaron effective mass to a realistic level, such that an Eliashberg treatment
\cite{eliashberg60,scalapino69,allen82,carbotte90,marsiglio08} makes sense. 

To more fully understand the effects due to NNN electron hopping we will first present our perturbation theory calculations for 
the energy and effective mass of the NNN Holstein model in one, two, and three dimensions. We use square and simple cubic
lattices for two and three dimensions, respectively. These results agree for sufficiently
low coupling strength with our 
exact numerical calculations using our previously refined algorithm for the Holstein model \cite{holsteinchandler14} extended to 
include NNN interactions. We note that quantitative agreement with perturbation theory extends over a surprisingly limited range
of electron-phonon interaction strength, even in three dimensions, which is the most applicable to bulk superconductors.
A low phonon frequency approximation to the perturbation theory results suggests a scaling of the phonon frequency with the
low energy effective bandwidth, which explains the results obtained as a function of NNN electron hopping. We also note an additional
scaling factor that accounts for the results with non-zero NNN hopping with respect to those with nearest-neighbour hopping only, over a more extended coupling strength range.

Since including NNN electron hopping also modifies the `effective' electronic bandwidth (to be defined more precisely below), we
should account for this in using the appropriate phonon frequency. That is, since altering the adiabatic ratio $\omega_E/t$, even
in the case with NN hopping only, is known to lead to changes in the polaronic effective mass for the same coupling strength, then
we should be careful to use an appropriately scaled phonon frequency.

After a brief introduction we use perturbation theory to determine the polaron effective mass in weak coupling. Since these
expressions are analytical, they are well-suited to examine the various scaling factors. We then present exact solutions, in
one, two, and three dimensions, to examine the effect on polaron mass over the entire coupling range.  We also note a
heuristic scaling, found numerically, that very accurately maps the parameters with NNN hopping back to those without,
before closing with a summary.\par

\bigskip
\medskip

\noindent {\bf MODEL \& METHODS}
\vskip1.0cm


{\it Holstein Model}
\vskip0.7cm

The Holstein model \cite{holstein1959} is perhaps the simplest model for describing electron-phonon interactions; it treats 
(optical) phonons as local ion vibrations, and assumes that each atomic site oscillates with the same characteristic frequency $\omega_E$.
With NNN hopping included, the Hamiltonian that describes such a system is:

\begin{align}
\hat{H} = &-t\sum_{\mathbf{j} , \mathbf{\delta} }(\hat{c}_{\mathbf{j}}^\dag\hat{c}_{\mathbf{j} + \mathbf{\delta}} + \hat{c}_{\mathbf{j} + \mathbf{\delta} }^\dag \hat{c}_{\mathbf{j}} ) -t_2\sum_{\mathbf{j},\mathbf{\gamma}} (\hat{c}_{\mathbf{j}}^\dag\hat{c}_{\mathbf{j}+\mathbf{\gamma}} + \hat{c}_{\mathbf{j}+\mathbf{\gamma}}^\dag \hat{c}_{\mathbf{j}} ) \nonumber   \\
			    &+\hbar\omega_E\sum_{\mathbf{j}}\hat{a}_{\mathbf{j}}^\dag\hat{a}_{\mathbf{j}} + \hbar\omega_Eg\sum_{\mathbf{j}}(\hat{a}_{\mathbf{j}}+\hat{a}_{\mathbf{j}}^\dag)\hat{c}_{\mathbf{j}}^\dag\hat{c}_{\mathbf{j}}.
\end{align}
Here, \(t\) and \(t_2\) are the nearest neighbour and NNN hopping integrals respectively with $\delta$ and $\gamma$ being the vectors to the 
nearest neighbour and next nearest neighbour sites, respectively. The sum over the vector of site positions $\mathbf{j}$ covers all sites.   
The electron creation and phonon creation operators at site $\mathbf{j}$ are  \(\hat{c}_{\mathbf{j}}^\dag\) and \(\hat{a}_{\mathbf{j}}^\dag\),
respectively, and 
\(g \equiv \sqrt{\frac{\alpha^2}{2\hbar M\omega_E^3}}\) is a dimensionless measure of the electron-phonon
coupling strength, with \(M\) being the atomic mass and \(\alpha\) being the coupling strength as defined in real space. 


In order to diagonalize this Hamiltonian, we transform into \(k\)-space, according to the equation:

\begin{equation}
\hat{c}_j = \frac{1}{\sqrt{N}}\sum_k e^{i\vec{k}\cdot\vec{R_j}}\hat{c}_k
\end{equation}
where \(\vec{R_j}\) points to lattice site \(j\), and \(\vec{k}\) is a wave vector summed over the First Brillouin Zone (FBZ). The relation for \(\hat{c}_j^\dag\) may be obtained simply by taking the Hermitian conjugate of the above expression, and the bosonic Fourier transforms are defined almost identically. In the FBZ there are \(N\) distinct \(k\) values within \((-\pi/a,\pi/a)\)
in each direction. The transformed Hamiltonian then becomes

\begin{align}\label{eq:hamiltonian}
\hat{H} = &\sum_k\epsilon (\vec{k})\hat{c}_k^\dag\hat{c}_k + \hbar\omega_E\sum_k\hat{a}_k^\dag\hat{a}_k \nonumber \\
			    &+\hbar g\omega_E\sum_{k,q}(\hat{a}_{q} + \hat{a}_{-q}^\dag)\hat{c}_{k+q}^\dag\hat{c}_{k},
\end{align}
and holds for all dimensions with dispersion relations \(\epsilon(\vec{k})\) given in \Cref{table1}.

\begin{table}\label{table1}
\bgroup
\def\arraystretch{1.9}
\begin{ruledtabular}
\begin{tabular}{ll}

Dim.    & $\epsilon (\vec{k})$ \\ \hline
1D & $-2t\cos{ka} -2t_2\cos{2ka} $\\
2D & $-2t(\cos{k_xa} + \cos{k_ya}) -4t_2\cos{k_xa}\cos{k_ya} $ \\
3D & $-2t(\cos{k_xa} + \cos{k_ya} + \cos{k_za}) $ \\
   & $\text{ }-4t_2(\cos{k_xa}\cos{k_ya} + \cos{k_xa}\cos{k_za} + \cos{k_ya}\cos{k_za})$\\

\end{tabular}
\end{ruledtabular}
\egroup
\caption{\small{Electron dispersion relations for the Holstein model, allowing for next-nearest neighbour hopping.}} 
\label{table1}
\end{table}

In this paper, we will examine various properties of the ground state, which for \(\frac{t'}{t} > -\frac{1}{4}\) is 
at zero total crystal momentum \(p\). For all dimensions this results in a low energy dispersion $E_p$
quadratic in \(p = \vert \mathbf{p} \vert\), 
so that the ground state effective mass of the electron is given by:

\begin{equation}\label{eq:mass}
\frac{1}{m^*} = \frac{1}{\hbar^2}\left.\frac{\partial^2E_p}{\partial p^2}\right\vert_{p=0}.
\end{equation}

\bigskip
\bigskip



{\it Weak-coupling Perturbation Theory}\par
\vskip0.75cm

Beginning with the perturbative approach (in the weak coupling regime), we consider the electron-phonon interaction to be the perturbation, so that the unperturbed energy is simply \(E_p^{(0)} = \epsilon(\vec{p}) \equiv \epsilon_p\). 
The unperturbed ground state for arbitrary \(p\) is therefore

\begin{equation} 
\label{eq:groundstate}
\ket{\phi_p^{(0)}} = \hat{c}_p^\dag\ket{0}.
\end{equation}
Here, \(\ket{0}\) is simply the electron-phonon vacuum state. Unperturbed excited states include all states with a single electron and any number of phonons
such that the total crystal momentum still adds to \(p\). It is easy to check that given these definitions, \(E_p^{(1)} = 0\), independent of the choice of total
crystal momentum. Under these conditions, the energy correction to second order (in \(\alpha\) or \(g\)) is:

\begin{equation}\label{eq:ep2}
E_p^{(2)} = \sum_{k,q}\frac{\left\vert \bra{\phi_p^{(0)}}\hat{V}_{pert}\hat{c}_k^\dag\hat{a}_q^\dag\ket{0}\right\vert^2}{\epsilon_p - (\epsilon_k + \hbar\omega_E)}
\end{equation}
In this case, \(\hat{V}_{pert}\) is the electron-phonon interaction term of the Hamiltonian in Eq.~\ref{eq:hamiltonian}. Note that only one phonon is considered in the
unperturbed excited states because \(\hat{V}_{pert}\) only creates (annihilates) one phonon. Upon evaluating this sum, we may apply 
Eq.~\ref{eq:mass} to find the effective
mass. In order to do so, we convert the sum in Eq.~\ref{eq:ep2} to an integral over \(k\)-values, since the thermodynamic limit ($N \rightarrow \infty$) implies a continuum of \(k\) values between \(-\pi/a\) and \(\pi/a\).

In the one-dimensional case, the corrected energy according to second order perturbation theory (in \(g\)) is

\begin{align}
E_p^{1D} = &-2t\left(\cos{(pa)} + \beta\cos{(2pa)}\right) \nonumber \\
				   -   &\frac{(\hbar g\omega_E)^2}{4W}\frac{1}{\sqrt{(\rfrac{\overline{\omega}_E}{2} - \beta + 1 + \cos{pa} + \beta\cos{2pa})b^2}} \nonumber \\
				   \times  & \left[\frac{4\beta - 1 + b}{\sqrt{3\beta + \rfrac{\overline{\omega}_E}{2} + \cos{pa} + \beta\cos{2pa} + b}} \nonumber \right.\\
		           -   &\left. \frac{4\beta - 1 - b}{\sqrt{3\beta + \rfrac{\overline{\omega}_E}{2} + \cos{pa} + \beta\cos{2pa} - b}}\right]
\end{align}
where we have defined dimensionless parameters \(\overline{\omega}_E \equiv \frac{\hbar\omega_E}{t}\) with \(b \equiv \sqrt{1 + 8\beta(\beta + \rfrac{\overline{\omega}_E}{2} + \cos{pa} + \beta\cos{2pa})}\) and $\beta \equiv t_2/t$. The above result, substituted into Eq.~(\ref{eq:mass}), gives an expression for the effective ground-state electron mass \(m^*\) at $p=0$:

\begin{align}
\left(\frac{m_b}{m^*}\right)_{\text{1D}} =&\text{ } 1 - \lambda\overline{\omega}_E\left\lbrace
		\frac{8\beta^2+12\beta+4\beta\overline{\omega}_E+\rfrac{1}{2}}{((4+\overline{\omega}_E)b^2)^{\rfrac{3}{2}}} \right. \nonumber \\
\times & \left.\left[\frac{4\beta - 1 + b}{\sqrt{8\beta + 2 + \overline{\omega}_E + 2b}} - \frac{4\beta - 1 - b}{\sqrt{8\beta + 2 + \overline{\omega}_E - 2b}}	\right] \right.\nonumber \\
+ &\left.\frac{1}{\sqrt{(4+\overline{\omega}_E)b^2}}\left[\frac{-2\beta}{b}\left(\frac{1}{\sqrt{8\beta+2+\overline{\omega}_E + 2b}}\right.\right.\right.\nonumber \\
+&\left.\left.\left.\frac{1}{\sqrt{8\beta +2 + \overline{\omega}_E - 2b}}\right)\right.\right. \nonumber \\
+ &\left.\left.\frac{4\beta - 1 + b}{\sqrt{2}(8\beta + 2 + \overline{\omega}_E + 2b)^{\rfrac{3}{2}}}\left(\frac{4\beta}{b}+1\right)\right.\right.\nonumber \\
+ &\left.\left.\frac{4\beta - 1 - b}{\sqrt{2}(8\beta + 2 + \overline{\omega}_E - 2b)^{\rfrac{3}{2}}}\left(\frac{4\beta}{b} - 1\right)\right]\right\rbrace
\label{eq:1dmass}
\end{align}
where \(b\) is evaluated at \(p=0\), and with \(\lambda \equiv \frac{2g^2\omega_E}{W}\), where $W$ is the electronic bandwidth. This definition for $\lambda$ applies for 3D as well, though in 2D, \(\lambda\equiv \frac{g^2\omega_E}{2\pi t}\) is used. This definition is preferred by the authors since it captures better the density of states for a single electron in the band, and only differs by an integral multiple of \(\pi/2\) anyway. Note that in the above equation, we have normalized by the inverse of the electron band mass (unperturbed effective mass):

\begin{equation}
\frac{1}{m_b} = \left.\frac{1}{\hbar^2}\frac{\partial^2\epsilon_p}{\partial p^2}\right\vert_{p=0} = \frac{(2t + 8t_2)a^2}{\hbar^2}.
\label{bare_mass}
\end{equation}

%

More generally, evaluating for the second-order energy correction in two or three dimensions proves tedious, and has a cumbersome,
unenlightening answer (as in \cref{eq:1dmass}). For these cases, we have also integrated the result numerically to check our
analytical  results. In the figures that follow, these are referred to as ``numerically integrated perturbation theory.''

\begin{figure}
\begin{center}
\includegraphics[height=2.8in,width=3.7in]{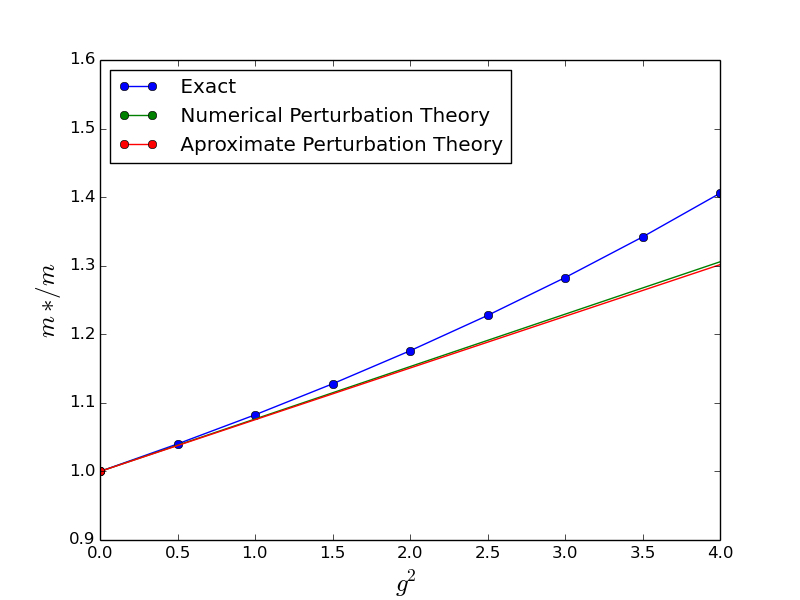}
\caption{  Numerically integrated perturbation theory, approximate perturbation theory, and exact numerical solution in 1D. Note that the 
approximate perturbation theory is on top of the numerically integrated perturbation theory so here the approximation in Table II
works very well. We
used parameter values of $\omega_E/t = 0.1 $ and $t_2/t = 0.025$. }
\label{fig:perturbation1D}
\end{center}
\end{figure}

\begin{figure}
\begin{center}
\includegraphics[height=2.8in,width=3.7in]{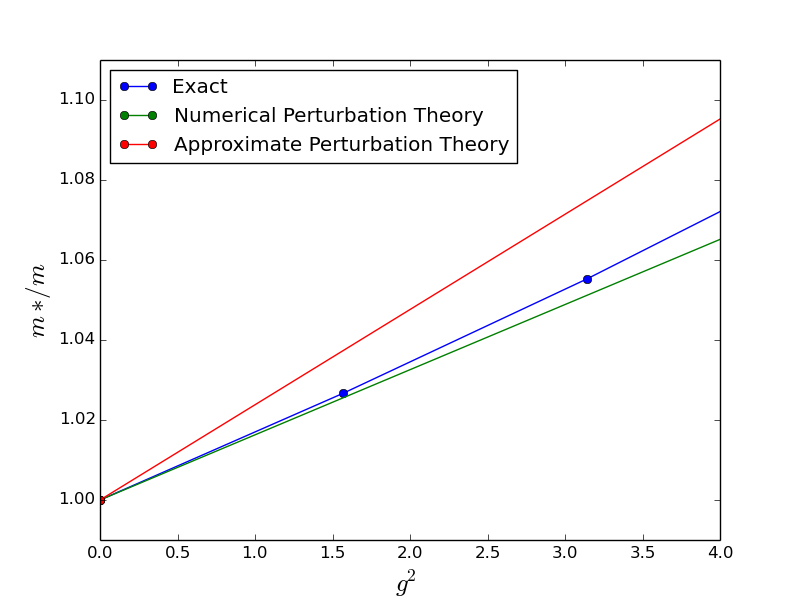}
\caption{  Numerically integrated perturbation theory, approximate perturbation theory, and exact numerical solution in 2D. Here the approximate 
perturbation calculation fails quite badly even in the very small perturbative regime from $g^2 = 0 $ to $g^2 = 1$.
We used parameter values of $\omega_E/t = 0.2 $ and $t_2/t = 0.025$. }
\label{fig:perturbation2D}
\end{center}
\end{figure}

\begin{figure}
\begin{center}
\includegraphics[height=2.8in,width=3.7in]{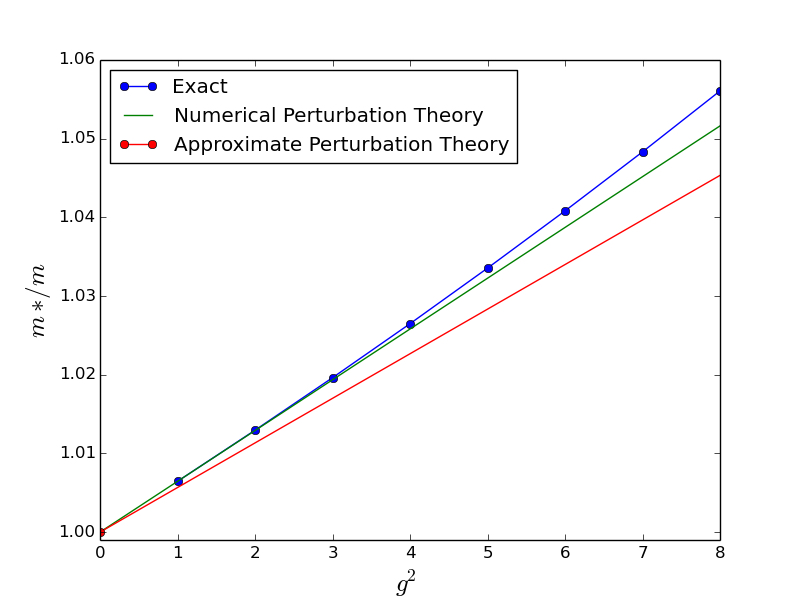}
\caption{  Numerically integrated perturbation theory, approximate perturbation theory, and exact numerical solution in 3D. Again the approximate 
perturbation calculation fails even in the very small perturbative regime from $g^2 = 0 $ to $g^2 = 3$.
We used parameter values of $\omega_E/t = 0.3$ and $t_2/t = 0.025$. }
\label{fig:perturbation3D}
\end{center}
\end{figure}

For an approximate analytical result, 
we observe that the integrand in the energy correction expression decays more or less to 0 by some
\(k_0 < \pi\) for most small values of the parameters \(\beta\) and \(\omega_E\). Cutting the integral off at this \(k_0\) and making the
approximation that \(k \ll 1\) for \(k \leq k_0\), we achieve the analytic approximations shown in \Cref{table2}. Unfortunately, these
approximations prove to be rather crude in 2D and 3D, which limits their usefulness. Regardless, we present them alongside 
our numerically integrated 
results for completeness in Figs.~(\ref{fig:perturbation1D} - \ref{fig:perturbation3D}). The approximations do work better for smaller $\omega_E$; however we have tested then in a reasonably physically representative regime of small $\omega_E$ \cite{holsteinchandler14} and even here the agreement is poor. 

Secondly, it is important to note that even without approximations, the range of coupling 
strengths over which the numerical perturbation theory is valid is very small. This feature is similar to our results for the standard 
Holstein model \cite{holsteinchandler14} so we do not recommend using the perturbation calculations for physical predictions 
but only as a check on more powerful
numerical calculations such as the Trugman Method.

\begin{table}
\bgroup
\def\arraystretch{1.9}
\begin{ruledtabular}
\begin{tabular}{lll}

Dim. & Eff. Mass Ratio:\hspace{2mm} $\displaystyle \frac{m_b}{m^*}$	 & $\displaystyle \frac{m_b}{m^*}$ (as a function of g) \\ \hline
1D & $\displaystyle 1 - \frac{1}{2}\frac{\lambda}{\sqrt{1+4\beta}}\frac{1}{\sqrt{\overline{\omega}_E}}$ & 
 $\displaystyle 1 - \sqrt{\frac{\overline{\omega}_E}{(1+4\beta)}}\frac{g^2}{4}$ \\ 
2D & $\displaystyle 1 -\frac{1}{2}\frac{\lambda}{(1+2\beta)}$ & $\displaystyle 1 -\frac{\overline{\omega}_E}{(1+2\beta)}\frac{g^2}{4\pi}$ \\ 
3D & $\displaystyle 1 - \frac{3}{4\pi}\frac{\lambda}{(1+4\beta)^{\rfrac{3}{2}}}\sqrt{\overline{\omega}_E}$ & 
$\displaystyle 1 -\frac{\overline{\omega}_E^{\rfrac{3}{2}}}{(1+4\beta)^{\rfrac{3}{2}}}\frac{g^2}{8\pi}$\\ 

\end{tabular}
\end{ruledtabular}
\egroup
\caption{\small{Approximate ground state effective mass for small 
$\bar{\omega}_E \equiv \omega_E/t$ and small $\beta \equiv t_2/t$, from weak-coupling perturbation theory.}}
\label{table2}
\end{table}

\bigskip
\bigskip


{\it Modified Trugman Method for Exact Numerical Solutions}\par
\vskip0.75cm

The single polaron problem is solved here with the variational exact diagonalization method described in Bon\u{c}a
et al. \cite{bonca99} and revised by the authors as described in \cite{li10,holsteinchandler14} to account for a rapidly 
growing Hilbert space from the additional terms in the Hamiltonian. For the data included in these plots 20 - 100 preliminary 
diagonalizations were performed with the most strongly contributing basis states selected at each iteration to seed the next 
iteration. All results were converged for the effective mass. While our algorithm for generating basis states differs significantly from that of
Chakraborty et al., \cite{chakraborty2011} our 1D results are in agreement with theirs and our algorithm permits efficient calculations in higher dimensions
and for realistic values of $\omega_E/t$. 

Note that the introduction of non-zero $t_2$ leaves the bandwidth invariant; however it does alter the curvature of the 
dispersion relation near $k=0$. The result is a different bare electron mass (see Eq.~(\ref{bare_mass})) and, 
correspondingly, a different \textit{effective} bandwidth. The different bare electron mass and different normalization 
of the effective mass has been noted before,\cite{chakraborty2011} but the different effective bandwidth was not considered.
If we take some small \(k_0 \ll 1/a\), and calculate
the bandwidth in the region \([-k_0,k_0]\) up to second order accuracy in \(k_0\), then we find the ratio of next-nearest neighbour bandwidth to nearest neighbour bandwidth to be:

\begin{align} \label{eq:bandwidth1}
&\left(\frac{W_{_{NNN}}}{W_{_{NN}}}\right)_{1D} \approx (1+4\beta) \approx  \left(\frac{W_{_{NNN}}}{W_{_{NN}}}\right)_{3D} \\
&\text{and:} \nonumber \\
\label{eq:bandwidth2}
&\left(\frac{W_{_{NNN}}}{W_{_{NN}}}\right)_{2D} \approx (1+2\beta).
\end{align}

This suggests the phonon energy scale $\hbar \omega_E$ also should be rescaled by the same factors to keep the ratio 
of phonon energy to effective bandwidth constant. 
The interaction strength parameter $g$ is dimensionless and thus remains unchanged, though 
the electron-phonon interaction term in the Hamiltonian is rescaled since it is proportional not simply to $g$, but to $ g \hbar \omega_E$.   
It can be seen from the result in \cref{table2}  that rescaling \(\overline{\omega}_E\) by the effective bandwidth change 
would transform the NNN approximate effective mass onto that for NN hopping only. In other words, if we use a renormalized
phonon frequency with the same value with respect to the effective electronic bandwidth, then the addition of next nearest
neighbour hopping has no effect on the effective mass (according to Table~\ref{table2}). However these are only approximate
perturbation calculations and the exact results show that the NNN effective mass is substantially different from the NN effective mass
even when the proper scalings have been taken into account. 

On the other hand, this rescaling of $\omega_E$ (which was not done by Chakraborty et al.
\cite{chakraborty2011}), definitely reduces the effect of the NNN hopping on the polaron effective mass.

\begin{figure}
\begin{center}
\includegraphics[height=2.8in,width=3.7in]{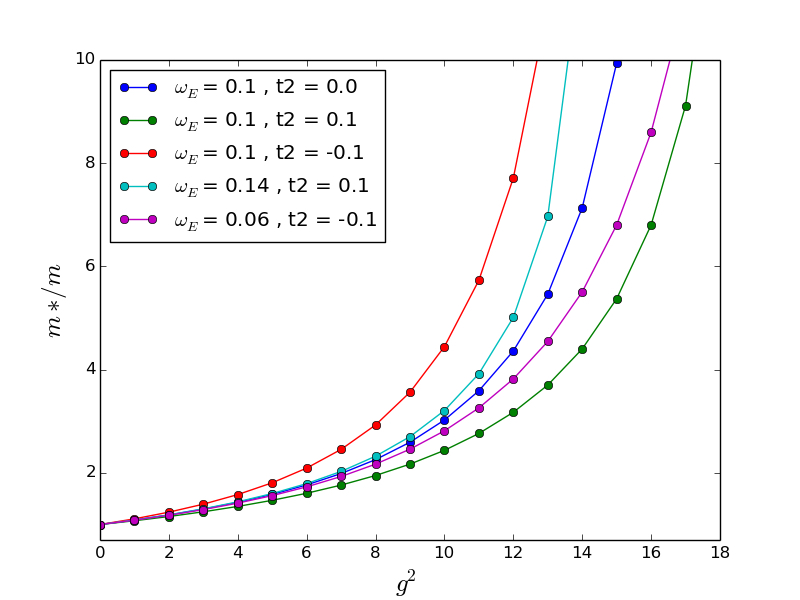}
\caption{  Exact numerical results in 1D for $t_2/t = 0, \pm 0.1$. Note that the effect of NNN hopping when the phonon frequency is scaled 
is smaller that that reported previously by Chakraborty et al. for identical values of NNN hopping. The scaling we have used also changes the direction of the scaling, as positive $t_2$ now raises the effective mass and negative $t_2$ lowers it. We use a relatively small
value of $\omega_E/t = 0.1$ since phonon energies near the adiabatic regime are representative of real phonons in real materials.}
\label{fig:exact1D}
\end{center}
\end{figure}

In Fig.~\ref{fig:exact1D} we compare the data with and without the scaling of $\omega_E$ to the original Holstein model in 
1D with $t_2/t = \pm 0.1$.  The 1D case has no crossover coupling strength in the standard Holstein model, where the polaron
effective mass suddenly begins to increase exponentially with coupling strength, and while $t_2 > 0$ did decrease the effective mass somewhat, no crossover point was found for the NNN Holstein model in 1D either. Our results are in agreement with Chakraborty et al. \cite{chakraborty2011} for the same value (i.e. unscaled) of $\omega_E$. However, including the correction
described in the preceding paragraph decreases the effective mass change and even the direction of the change. The 1D case is not
very realistic for bulk materials so we continue with 2D and 3D calculations.

\begin{figure}
\begin{center}
\includegraphics[height=2.8in,width=3.7in]{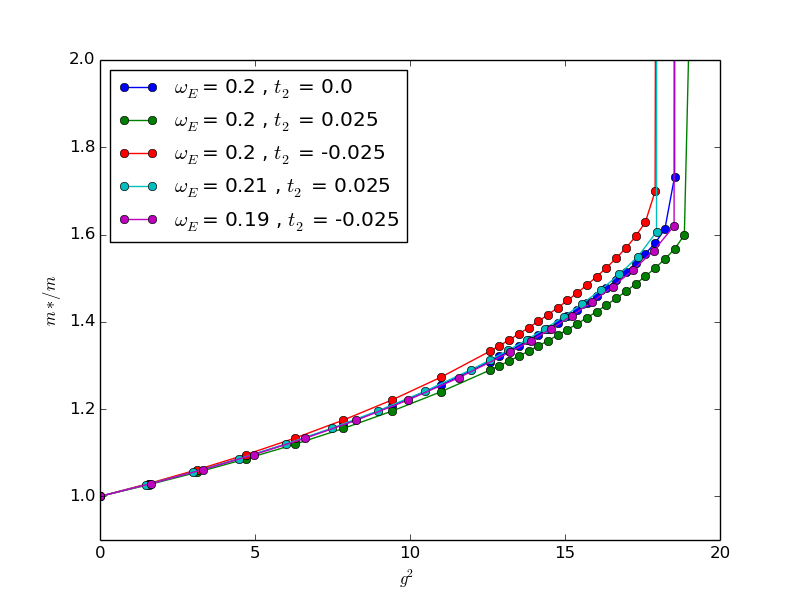}
\caption{  Exact numerical results in 2D for $t_2/t = 0, \pm 0.025$. Note that the scaling of $\omega_E$ maps the NNN calculations back to the original.
This is in agreement with our approximate perturbation theory results, though perhaps serendipitously since the approximate perturbation theory was not 
very close to the exact numerical perturbation theory. }
\label{fig:exact2D}
\end{center}
\end{figure}

\begin{figure}
\begin{center}
\includegraphics[height=2.8in,width=3.7in]{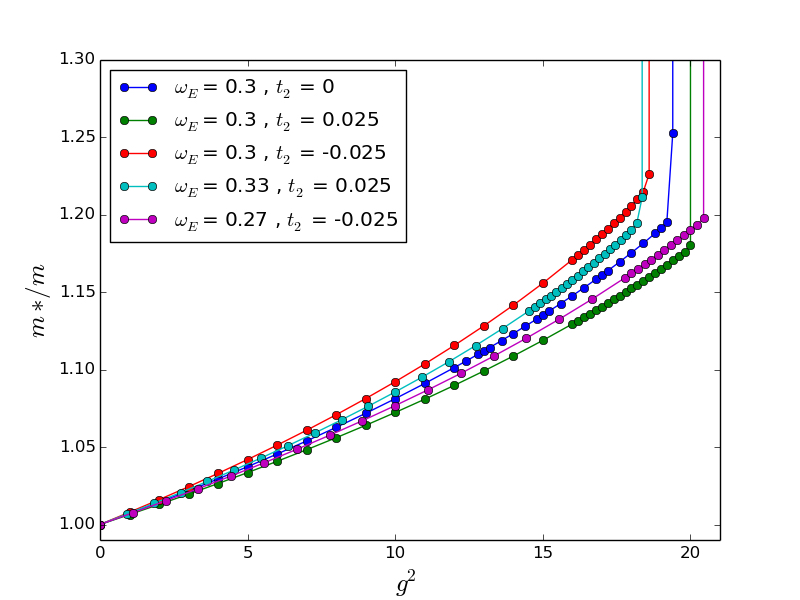}
\caption{  Exact numerical results in 3D for $t_2/t =  0, \pm 0.025$. Note the change in the crossover coupling strength. Here it seems that the rescaling of
$\omega_E$ \textit{enhances} the effect of the NNN hopping on the effective mass beyond the crossover point, albeit reversed in sign as was the case in 1D. }
\label{fig:exact3D}
\end{center}
\end{figure}

In two and three dimensions we find that for small NNN hopping parameters the effective mass deviates from the standard Holstein model only slightly until the crossover coupling strength in 2D and 3D is reached. At this point the effective masses increases sharply for both models. However, the introduction of non-zero $t_2$ changes the crossover point slightly. We have included the unscaled results with the 
$\omega_E/t$ scaled results
for the sake of completeness in Fig.~\ref{fig:exact2D} and Fig.~\ref{fig:exact3D} for 2D and 3D, respectively.

While our approximate perturbation theory suggests that the scalings of $\omega_E/t$ given in Table~\ref{table2} would map the NNN effective masses onto the NN effective mass, Figs.~\ref{fig:exact1D}, \ref{fig:exact2D}, and \ref{fig:exact3D} make it clear that this
scaling does not work very well except in the 2D case. The approximate perturbation theory only agrees exactly with the numerical perturbation theory in 1D and there is no reason
to trust even the numerical perturbation theory for the effective mass outside of very small coupling strength as we have remarked on other 
occasions.\cite{holsteinchandler14} In 2D the approximate $\omega_E$ scaling suggested in Table~\ref{table2} mapped the effective masses back close to that obtained with no NNN hopping, but in 1D and 3D it 
over-corrected the change in the effective mass. Also, not surprisingly, in all three dimensions a simple scaling does not work well in the 
very strong coupling regime. So in agreement with Chakraborty et al., \cite{chakraborty2011} NNN hopping does introduce changes in the properties of the polaron, and leads to a decreased effective mass for some additional (positive) $t_2$ hopping if no 
phonon frequency scaling
is introduced. But NNN hopping leads to an {\it increased} effective mass if the phonon frequency is also increased to account for the increase in the `local'
bandwidth. Most importantly for our understanding of the conventional framework for superconductivity, 
the inclusion of NNN hopping changes the critical
coupling strength in 3D at which the effective mass increases sharply towards infinity (see the large coupling regime of 
Fig.~\ref{fig:exact3D}).

\bigskip
\bigskip


{\it Heuristic Scaling}\par
\vskip0.75cm

\begin{figure}
\begin{center}
\includegraphics[height=2.8in,width=3.7in]{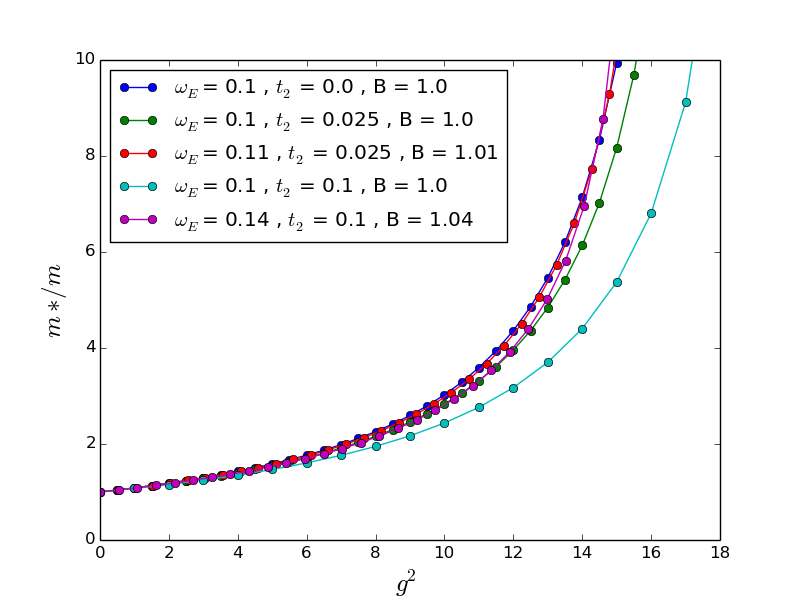}
\caption{  Exact numerical results in 1D, with g scaling for $t_2/t = 0, \pm 0.1$. Note that the scale is much larger than that of Fig. (\ref{fig:exact1D}) and the scaling works very well over the previous range, and improves the agreement with the numerical
results at stronger couplings.}
\label{fig:omegaEandgscaled1D}
\end{center}
\end{figure}

\begin{figure}
\begin{center}
\includegraphics[height=2.8in,width=3.7in]{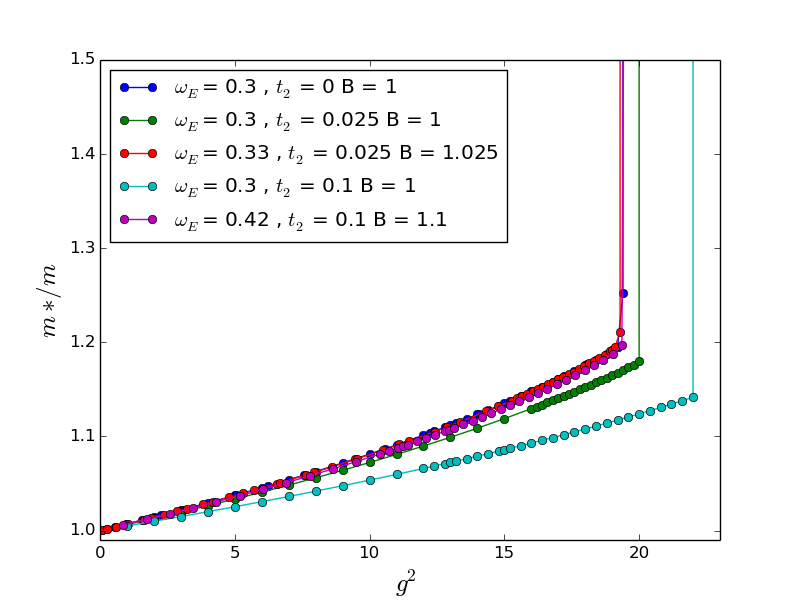}
\caption{  Exact numerical results in 3D , with g scaling for $t_2/t = 0, 0.025$, and $0.1$. The scaling works remarkably well
over the entire range, and matches up the crossover points very well.}
\label{fig:omegaEandgscaled3D}
\end{center}
\end{figure}

In the course of our investigations we further found a heuristic scaling of the coupling strength that, combined with the bandwidth-inspired scaling, works very well; however, the underlying physical motivation is still rather unclear. We introduce a scaling factor in the dimensionless interaction parameter $ g_{original} \cdot B = g_{scaled}$ such that the new term in the Hamiltonian is: 
\begin{equation}
 \hbar\omega_E g_{scaled} \sum_{\mathbf{j}}(\hat{a}_{\mathbf{j}}+\hat{a}_{\mathbf{j}}^\dag)\hat{c}_{\mathbf{j}}^\dag\hat{c}_{\mathbf{j}}
\end{equation}
By experimenting with different values of $B$ we found that in 1D, $B = 1 + 4\beta $ and in 3D, $B = 1 + \beta$ work very well. These
statements are validated in Figs.~(\ref{fig:omegaEandgscaled1D} , \ref{fig:omegaEandgscaled3D}) and at least for 3D is very accurate,
even in the crossover regime.

\bigskip
\bigskip


\noindent{\bf SUMMARY AND CONCLUSIONS}
\vskip1.0cm

We have performed weak coupling perturbation calculations for the Holstein model with next nearest neighbour (NNN) hopping and compared them with exact calculations in one, two, and three dimensions. We confirmed previous results obtained in one dimension
by Chakraborty et al.; \cite{chakraborty2011} however, we point out that a more appropriate comparison, at least in weak coupling,
requires a change of the phonon frequency, $\omega_E$, as given in Table~\ref{table2}, for the case with NNN hopping. With this 
change accounted for we find that the effect of NNN hopping on the effective mass is opposite to the change without a phonon frequency
change. Including $t_2$ with the same sign as $t$ reduces the polaron effective mass, when the same phonon frequency is used,
whereas including a change in the phonon frequency according to the changes in effective bandwidth as in Eqs.~(\ref{eq:bandwidth1})
and (\ref{eq:bandwidth2}) {\it increases} the effective mass in 1D and 3D. We feel that including the change in phonon frequency is
the more physically correct procedure. In two
dimensions NNN hopping has very little effect on the effective mass in weak coupling. 

In 1D and 3D we have also found a heuristic scaling factor for $g$, the dimensionless electron-phonon coupling strength, that maps the results for the polaron effective mass of the NNN Holstein model back onto the standard model without NNN hopping. While the physical
reason for this effect is not known, this heuristic scaling allows us to crudely 
estimate how increasing $t_2$ impacts the coupling strength at which the (sharp) crossover occurs for polaronic behaviour.
For small values of $t_2/t$ the crossover remains in the regime of moderate electron-phonon coupling. Therefore it remains difficult to reconcile the fairly strong coupling attributed to some real metals/superconductors with the diverging effective mass predicted for
a single polaron at the same coupling strength.

\bigskip

\noindent {\bf ACKNOWLEDGEMENTS}

This work was supported in part by the Natural Sciences and Engineering
Research Council of Canada (NSERC). In addition, this work was made possible in part by an NSERC USRA to Christian Prosko and 
an Alberta Ingenuity Len Bolger Scholarship to Carl Chandler.

\bigskip

\noindent{\bf Competing Financial Interests}

The authors declare no competing financial interests.

\bigskip

\noindent $^\ast$Correspondence to fm3@ualberta.ca

\bigskip


\bigskip

\noindent {\bf Author Contributions}

CJC and FM originally formulated the problem addressed in this paper. All authors contributed to the calculations --- CJC performed the exact calculations and CP and FM did the perturbation calculations.
All authors contributed significantly to the writing of the manuscript.



\begin{thebibliography}{99}

\bibitem{bardeen1957} Bardeen, J., Cooper, L.N., and Schrieffer, J.R. {\rm Theory of Superconductivity}. \href{http://journals.aps.org/pr/pdf/10.1103/PhysRev.108.1175}{{\it Phys. Rev.} {\bf 108}, 1175-1204, (1957)}.
\bibitem{alexandrov2011} Alexandrov, A.S. {\rm High-temperature superconductivity: the explanation}. \href{http://iopscience.iop.org/article/10.1088/0031-8949/83/03/038301/pdf}{{\it Physica Scripta}, {\bf 83}, 038301-1-8 (2011)}.

\bibitem{hirsch15} Hirsch, J.E. et al. {\it Special Issue on Superconducting Materials (eds Hirsch, J.E. et al.)} 
\href{http://www.sciencedirect.com/science/journal/09214534/514/supp/C}{{\it Physica C} {\bf 514}, 1-443 (2015).} 

\bibitem{alexandrov07} Alexandrov S.A. et al. {\it Polarons in Advanced
Materials, Vol. 103 (ed  Alexandrov, A.S.)} {\it Springer Series in Material Sciences, Springer Verlag, Dordrecht (2007).}

\bibitem{bonca99} Bon\u{c}a J., Trugman, S.A., and Batist\'{\i}c, I. {\rm Holstein polaron}. \href{http://journals.aps.org/prb/pdf/10.1103/PhysRevB.60.1633}{{\it Physical Review B} {\bf 60}. 1633-1642 (1999)}.

\bibitem{li10} Li, Z., Baillie, D., Blois, C., and Marsiglio, F. {\rm Ground-state properties of the Holstein model near the adiabatic limit.}
\href{http://journals.aps.org/prb/pdf/10.1103/PhysRevB.81.115114}{{\it Phys. Rev.} B{\bf81}, 115114-1-6 (2010)}.

\bibitem{alvermann10} Alvermann, A., Fehske, H., and Trugman, S.A. {\rm Polarons and slow quantum phonons.} \href{http://journals.aps.org/prb/pdf/10.1103/PhysRevB.81.165113}{{\it Phys. Rev.} B{\bf 81}, 165113-1-9 (2010).}

\bibitem{ku02} Ku, L-C., Trugman, S.A., and Bon\v ca, J. {\rm Dimensionality Effects on the Holstein polaron}. \href{http://journals.aps.org/prb/pdf/10.1103/PhysRevB.65.174306}{{\it Phys. Rev.} B\textbf{65}, 174306-1-10 (2002).}

\bibitem{li12} Li, Z., and Marsiglio, F.  {\rm The Polaron-Like Nature of an Electron Coupled to Phonons.}
\href{http://download.springer.com/static/pdf/932/art%253A10.1007%252Fs10948-012-1601-6.pdf?originUrl=http%3A%2F%2Flink.springer.com%2Farticle%2F10.1007%2Fs10948-012-1601-6&token2=exp=1465328756~acl=%2Fstatic%2Fpdf%2F932%2Fart%25253A10.1007%25252Fs10948-012-1601-6.pdf%3ForiginUrl%3Dhttp%253A%252F%252Flink.springer.com%252Farticle%252F10.1007%252Fs10948-012-1601-6*~hmac=71d3191fc10e644bc9bc492adfb9a01cdf39968fbb56965f6a9570a7b27f7f26}{{\it J. Supercond. Nov. Magn.} {\bf25}, 1313-1317 (2012).}

\bibitem{davenport12} Davenport, A.R., Hague, J.P., and Kornilovitch, P.E. {\rm Mobile small bipolarons on a three-dimensional cubic lattice.} \href{http://journals.aps.org/prb/pdf/10.1103/PhysRevB.86.035106}{{\it Phys. Rev.} B{\bf 86}, 035106-1-11 (2012).}

\bibitem{bonca01} Bon\v ca, J. and  Trugman, S.A. {\rm Bipolarons in the extended Holstein Hubbard model.} \href{http://journals.aps.org/prb/pdf/10.1103/PhysRevB.64.094507}{{\it Phys. Rev.} B{\bf 64}, 094507-1-4 (2001).}

\bibitem{holsteinchandler14} Chandler, C.J., and Marsiglio, F. {\rm Extended versus standard Holstein model: Results in two and three dimensions.} \href{http://journals.aps.org/prb/pdf/10.1103/PhysRevB.90.125131}{{\it Phys. Rev.} B{\bf 90}, 125131-1-8 (2014).}

\bibitem{chakraborty2011} Chakraborty, M.,  Das, A. N., and Chakrabarti, A. {\rm Study of the one-dimensional Holstein model with next-nearest-neighbor hopping.} \href{http://iopscience.iop.org/article/10.1088/0953-8984/23/2/025601/pdf}{{\it Journal of Physics: Condensed Matter}, {\bf 23}, 025601-1-7 (2011).}

\bibitem{gladstone69} Gladstone, G.,  Jensen, M.A., and
Schrieffer, J.R. {\it Superconductivity in the Transition Metals: Theory and Experiment} in {\it Superconductivity}, {\it ed Parks, R.D.,}
(Marcel Dekker, Inc., New York, 1969)p. 665-816. 

\bibitem{eliashberg60} Eliashberg, G.M., {\it Zh. Eksp. Teor.Fiz.} {\bf 38} 966-976 (1960); {\it Sov. Phys. JETP} {\bf 11} 696-702 (1960).

\bibitem{scalapino69} Scalapino, D.J., {\it The Electron-phonon Interaction and Strong-coupling Superconductors}, in 
{\it Superconductivity}, {\it ed Parks, R.D.,} (Marcel Dekker,New York, 1969), p. 449-560.

\bibitem{allen82} Allen, P.B., and Mitrovi\'c, B., {\it Solid State Physics Vol. 37}, {\it eds Ehrenreich, H., Seitz, F., and Turnbull, D.}, 
(Academic, New York, 1982), p. 1-92.

\bibitem{carbotte90} Carbotte, J.P., {\rm Properties of boson-exchange superconductors.} \href{http://journals.aps.org/rmp/pdf/10.1103/RevModPhys.62.1027}{{\it Rev. Mod. Phys}. {\bf 62} 1027-1157 (1990).}

\bibitem{marsiglio08} Marsiglio, F. and Carbotte, J.P., `\href{http://arxiv.org/pdf/cond-mat/0106143v1.pdf}{Electron-Phonon Superconductivity}', in {\it Superconductivity, Conventional and Unconventional Superconductors},
{\it eds Bennemann K.H. and Ketterson J.B.}, (Springer-Verlag, Berlin, 2008), pp. 73-162.

\bibitem{holstein1959} Holstein, T., {\rm Studies of polaron motion : Part I. The
molecular-crystal model.} \href{http://ac.els-cdn.com/0003491659900028/1-s2.0-0003491659900028-main.pdf?_tid=62dfc1d0-2ce9-11e6-9021-00000aab0f02&acdnat=1465329337_b13719d5502be8e180491e0a31e47071}{{\it Annals of Physics} {\bf 8}, 325-342 (1959).}

\end{thebibliography}

\end{document}